\documentclass[preprint,12pt,pre]{revtex4}%
\usepackage{amsfonts}
\usepackage{amsmath}
\usepackage{amssymb}
\usepackage{graphicx}%
\providecommand{\U}[1]{\protect\rule{.1in}{.1in}}

\begin{document}
\preprint{UATP/0905}
\title{Comment on "Non-equilibrium entropy of glasses formed by continuous
cooling"\ [J. Non-Cryst. Solids \textbf{355} (2009) 600]}
\author{P. D. Gujrati}
\affiliation{Departments of Physics and Polymer Science, The University of Akron, Akron, OH 44325}

\begin{abstract}
We use the general statement of the second law applied to an isolated system,
the glass in an extremely large medium, to prove that the entropy of the glass
must decrease with time during its relaxation towards the supercooled liquid
state. This result contradicts the claim of Mauro et al \cite{Mauro} and their
computational result of the entropy, according to which their entropy of the
glass increases. Our approach using the isolated system completely bypasses
the issue of ergodicity loss in glasses, as discussed in the comment. 

\end{abstract}
\date[September 2, 2009]{}
\maketitle

Mauro et al \cite{Mauro} say that when a glass (GL) is held at a fixed
temperature \cite{Note0} for a period longer than the experimental time scale
$\tau$ used for its preparation, it tends spontaneously towards the
equilibrium state of the supercooled liquid (SCL), the process known as the
structural relaxation. During this process, they say that "the properties of a
glass, including entropy, slowly approach their equilibrium values." No one
can disagree with this statement. However, they also claim that the entropy
increases during this relaxation process. This they say follows from the
process being spontaneous. They propose a definition of the entropy for a
glass, which is then used to support the claim. This demonstration means that
their defined quantity most probably has the desired behavior, but leaves open
the question:\ Is this the correct definition of the entropy for a glass?
Mauro et al \cite{Mauro} certainly provide motivation for their definition,
but it is not obvious if their choice gives the correct entropy. This is by no
means a trivial issue and needs to be clarified. In this comment, we establish
on very general grounds that the claim is unjustified, which then casts doubt
on their method to calculate the entropy of the glass.

It is well known that when a liquid is disturbed suddenly from its equilibrium
state by changing the temperature or pressure of the medium (which contains
the liquid), then the liquid undergoes a rapid, solidlike change, followed by
a slower, liquidlike change towards the new equilibrium state characterized by
the medium. These changes can be seen in the variation in its thermodynamic
properties such as the volume $V$ or its enthalpy $H$ with time. For a SCL,
the above scenario plays an important role. As the temperature is lowered or
the pressure is increased, the scale separation between the fast and the slow
processes in SCLs increases until the latter becomes too large compared to the
experimental time $\tau$. In this case, the system is said to be kinetically
arrested in that the liquidlike changes no longer contribute to the observed
properties over a period close to $\tau$. The system behaves like a solid and
is called a \emph{glass}.

The glass is thus a system far from equilibrium so one cannot apply
equilibrium statistical mechanics or equilibrium thermodynamics to investigate
its properties, which vary with time, a point also made by Mauro et al
\cite{Mauro}. One must resort to apply non-equilibrium thermodynamics, not a
well-developed field at present, to study glasses and their relaxation in
time. Another possibility is to proceed in a general manner by following the
consequences of the second law applied to an isolated system (consisting of
the liquid and the medium), which is well established. The law is independent
of the details of the systems considered and does not requires any other
sophisticated concepts like ergodicity or its loss, etc. Most importantly, it
is not affected by the controversial issues raised by Mauro et al \cite{Mauro}
to study glass transition. In particular, the entropy, however one may wish to
define it, must satisfy the second law. If it does not, it is not the correct
entropy. As we will see, we do not have to worry about the converse: is the
entropy the correct entropy if it satisfies the second law?

According to the second law, the entropy $S_{0}$ of an isolated system
$\Sigma_{0}$ can never decrease in time:%
\begin{equation}
\frac{dS_{0}(t)}{dt}\geq0.\label{Second_Law}%
\end{equation}
What happens inside the isolated system (loss of ergodicity in parts of the
system, chemical reactions, phase changes, etc.) cannot affect the direction
of the inequality, which makes it the most general principle of
non-equilibrium thermodynamics. The law itself imposes no restriction on the
actual rate of entropy change, but this will not be relevant in our discussion
here. In general, $S_{0}$ also depends on the number of particles $N_{0}$,
energy $E_{0}$, and volume $V_{0}$ of $\Sigma_{0}$. Thus, $S_{0}(t)$ used
above should be really written as $S_{0}(E_{0},V_{0},N_{0},t)$. However, as
the extensive quantities remain constant in time there is no harm in using the
compact form $S_{0}(t)$ during approach to equilibrium.\ 

The equality in (\ref{Second_Law}) occurs when the isolated system is in
equilibrium so that the entropy $S_{0}(E_{0},V_{0},N_{0},t)$ has achieved its
maximum possible value and no longer has any explicit time-dependence and can
be simply written as $S_{0}(E_{0},V_{0},N_{0})$ or $S_{0}$. In this case,
different parts of $\Sigma_{0}$\ have the same temperature $T_{0}$ and
pressure $P_{0}$ determined by:
\begin{equation}
\frac{1}{T_{0}}=\frac{\partial S_{0}}{\partial E_{0}},\text{\ \ \ }\frac
{P_{0}}{T_{0}}=\frac{\partial S_{0}}{\partial V_{0}}.\label{Eq_Conds_0}%
\end{equation}
\ Otherwise, they have different temperatures and pressures, in which case a
common assumption made by almost all workers is that of partial equilibrium
(see, for example, Landau and Lifshitz \cite[see p. 13]{Landau}) when
$\Sigma_{0}$ is out of equilibrium; each part is in internal equilibrium
(local equilibrium) so that we can define the temperature, pressure, etc. for
each part. In this situation, their entropies have no explicit $t$-dependence
[see the equilibrium condition (\ref{Eq_Conds_0}) above for $\Sigma_{0}$];
their variation in times comes from the time variation of their energies,
volumes, etc. As said above, we will think of $\Sigma_{0}$\ consisting of only
two parts, the system of interest $\Sigma$ (SCL/GL) and the medium denoted by
$\widetilde{\Sigma}$ surrounding. The energy, volume and the number of
particles of $\Sigma$\ are denoted by $E$, $V$, and $N,$ respectively, while
that of the medium $\widetilde{\Sigma}$ by $\widetilde{E}$, $\widetilde{V}$,
and $\widetilde{N}.$ The medium is considered to be very large compared to
$\Sigma$. The entropy $S$ of the system in the glassy state, in which the fast
dynamics has equilibrated so that it is treated as in internal equilibrium,
determines its temperature $T(t)$ and $P(t)$:%
\begin{equation}
\frac{\partial S}{\partial E}=\frac{1}{T(t)},\ \frac{\partial S}{\partial
V}=\frac{P(t)}{T(t)}.\label{Eq_Conds}%
\end{equation}
These are standard relations for the entropy \cite{Landau}, except that all
quantities except $S$ may have an \emph{explicit} dependence on time $t$ that
will make $S$ depend \emph{implicitly} on time. Relations like these for
internal equilibrium are used commonly in non-equilibrium thermodynamics. For
example, we use them to establish that heat flows from a hot body to a cold
body; see Sect. 9 in Landau and Lifshitz \cite{Landau}. In the following, the
glass is considered to be formed under isobaric conditions, so that we will
take its pressure $P(t)$ to be always equal to $P_{0}$ of the medium, but its
temperature will in general be different than $T_{0}$ and vary in time.

Below the glass transition at $T_{\text{g}}$, GL ($\Sigma$) will relax so as
to come to equilibrium with the medium if we wait longer than $\tau$. It
should be obvious that the medium also has to be in internal equilibrium,
except that it is so large compared to the system that its temperature,
pressure, etc. are not affected by the system. Obviously,%
\[
E_{0}=E+\widetilde{E},\ \ V_{0}=V+\widetilde{V},\ \ N_{0}=N+\widetilde{N}.
\]
We will assume that $N$ of the system is also fixed, which means that
$\widetilde{N}$ is also fixed. However, the energy and volume of the system
change with $t$. The entropy $S_{0}$ of the isolated system can be written as
the sum of the entropies $S$ of the system and $\widetilde{S}$ of the medium:%
\[
S_{0}(E_{0},V_{0},N_{0},t)=S(E,V,N)+\widetilde{S}(\widetilde{E},\widetilde
{V},\widetilde{N});
\]
there is no explicit $t$-dependence on the right due to internal equilibrium.
The correction to this entropy due to the weak stochastic interactions between
the system and the medium has been neglected, which is a common practice. We
expand $S_{0}$ in terms of the small quantities of the system \cite{Landau}%
\[
\widetilde{S}(\widetilde{E},\widetilde{V},\widetilde{N})\simeq\widetilde
{S}(E_{0},V_{0},\widetilde{N})-\left.  \left(  \frac{\partial\widetilde{S}%
}{\partial\widetilde{E}}\right)  \right\vert _{E_{0}}E(t)-\left.  \left(
\frac{\partial\widetilde{S}}{\partial\widetilde{V}}\right)  \right\vert
_{V_{0}}V(t).
\]
It follows from the internal equilibrium of $\widetilde{\Sigma}$ and the
smallness of $\Sigma$\ that%
\[
\left.  \left(  \frac{\partial\widetilde{S}}{\partial\widetilde{E}}\right)
\right\vert _{E_{0}}=\frac{1}{T_{0}},\ \ \left.  \left(  \frac{\partial
\widetilde{S}}{\partial\widetilde{V}}\right)  \right\vert _{V_{0}}=\frac
{P_{0}}{T_{0}},
\]
see (\ref{Eq_Conds_0}), and $\widetilde{S}(E_{0},V_{0},\widetilde{N}),$ which
is a constant, is independent of the system. Thus,%
\begin{align}
S_{0}(E_{0},V_{0},N_{0},t)-\widetilde{S}(E_{0},V_{0},\widetilde{N}) &  \simeq
S(E,V,N)-E(t)/T_{0}-P_{0}V(t)/T_{0},\nonumber\\
&  =S(t)-H(t)/T_{0}=-G(t)/T_{0},\label{Gibbs_Free_Energy}%
\end{align}
where
\[
G(t)\equiv H(t)-T_{0}S(t),\ H(t)\equiv E(t)+PV(t)
\]
are the time-dependent Gibbs\ free energy and enthalpy of the system $\Sigma$
with the medium $\widetilde{\Sigma}$ at fixed $T_{0}$ and $P_{0}$. It should
be stressed that the $t$-dependence in $S,H$, and $G$ is implicit through
$E(t),$ and $V(t)$.

\qquad Let us consider the time derivative of the entropy $S_{0}$, which is
changing because the energy and volume of $\Sigma$ are changing with time
\cite{Landau}. Thus,%
\begin{align*}
\frac{dS_{0}(t)}{dt} &  =\frac{dS}{dt}-\frac{1}{T_{0}}\frac{dE(t)}{dt}%
-\frac{P_{0}}{T_{0}}\frac{dV(t)}{dt}\\
&  =\left(  \frac{\partial S}{\partial E}-\frac{1}{T_{0}}\right)  \frac
{dE(t)}{dt}+\left(  \frac{\partial S}{\partial V}-\frac{P_{0}}{T_{0}}\right)
\frac{dV(t)}{dt}\geq0,
\end{align*}
as the relaxation goes on in the system $\Sigma$. It is clear that
\[
\frac{\partial S}{\partial E}\neq\frac{1}{T_{0}},\ \frac{\partial S}{\partial
V}\neq\frac{P_{0}}{T_{0}},
\]
if $dS_{0}/dt>0$. Thus, as long as the relaxation is going on due to the
absence of equilibrium, the two inequalities must hold true. In a cooling
experiment, we expect the system $\Sigma$ to lose energy, so that $dE/dt<0.$
Accordingly, the derivative $\partial S/\partial E,$ which by definition
\ represents the inverse temperature $1/T(t)$ of the system \cite{Landau},
must be less than $1/T_{0}$ of the medium. In other words, $T(t)>T_{0}$ during
relaxation and approaches $T_{0}$ from above as the relaxation ceases when
equilibrium has achieved:%
\[
T(t)\geq T_{0}.
\]
As $\partial S/\partial V=P_{0}/T(t)$, we see immediately that
\begin{equation}
\frac{dS_{0}(t)}{dt}=\left(  \frac{1}{T(t)}-\frac{1}{T_{0}}\right)
\frac{dH(t)}{dt}\geq0,\label{Total_Entropy_Rate}%
\end{equation}
from which it follows immediately that $dH(t)/dt<0,$ the equality occurring
only when equilibrium has been achieved.%
\begin{figure}
[ptb]
\begin{center}
\includegraphics[
trim=1.001315in 7.225579in 2.790484in 0.647003in,
height=2.7968in,
width=4.3837in
]%
{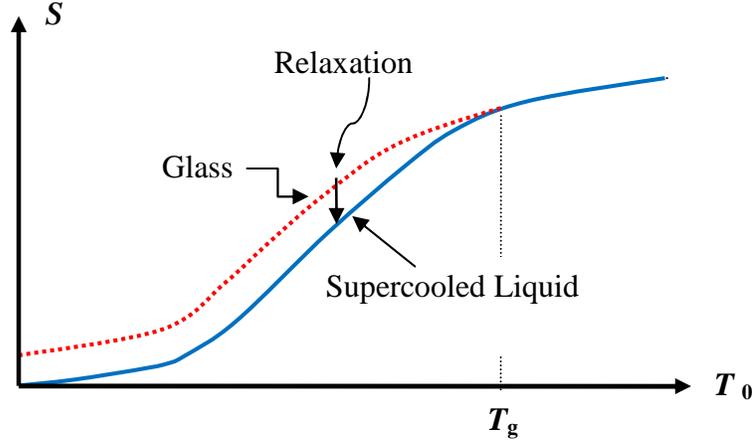}%
\caption{Schematic behavior of the entropy for SCL (blue curve) and GL\ (red
dotted curve). The GL entropy decreases, shown by the downward arrow, as it
isothermally (constant temperature $T_{0}$ of the medium) relaxes towards SCL,
during which its temperature $T(t)$ also decreases towards $T_{0}$. }%
\label{Fig_entropyglass}%
\end{center}
\end{figure}

The above calculation also shows that
\begin{equation}
\frac{dS(t)}{dt}=\frac{1}{T(t)}\frac{dH(t)}{dt},\label{Entropy_variation1}%
\end{equation}
which is the first term in (\ref{Total_Entropy_Rate}).

The relaxation that occurs in the glass originates from its tendency to come
to thermal equilibrium during which its temperature $T(t)$ varies with time;
recall that we are considering a cooling experiment. The relaxation process
results in the lowering of the corresponding Gibbs free energy, as is seen
from using (\ref{Gibbs_Free_Energy}) in (\ref{Second_Law}). Accordingly, there
are changes in its enthalpy and entropy, which are in the same direction; see
(\ref{Entropy_variation1}). The lowering of $G(t)$ with time results in not
only lowering the enthalpy in a cooling experiment, as observed experimentally
and demonstrated above, but also the entropy $S(t)$ during relaxation:%
\begin{equation}
(dS(t)/dt)\leq0,\label{Entropy_variation}%
\end{equation}
as shown in Fig. \ref{Fig_entropyglass}. We will now suppress the
$t$-dependence in $S$ for simplicity. As the entropy of GL decreases as it
relaxes towards SCL, we have%
\begin{equation}
S_{\text{SCL}}=S_{\text{GL}}+S_{\text{relax}},\label{Relax_S_Part}%
\end{equation}
where
\begin{equation}
S_{\text{relax}}<0\label{Relax_S}%
\end{equation}
below $T_{\text{g}}$. The inequality (\ref{Relax_S}) follows from the second
law, and cannot be violated$.$ The lowering of the glass entropy
$S_{\text{GL}}$ is not a violation of the second law, whose statement in the
form of the law of increase of entropy is valid only for an isolated system.
When the system is in contact with a medium, it is the corresponding free
energy that decreases with time. This is seen clearly from
(\ref{Gibbs_Free_Energy}), which shows that it is the difference
$S_{0}(t)-\widetilde{S}=-G(t)/T_{0}$ that increases with time, or the Gibbs
free energy $G(t)$ decreases with time. The relaxation of the system in the
present case, although spontaneous, does not mean that its entropy must increase.

We now need to turn our attention to the configurational entropy
$S^{\text{(c)}}$, which is defined as that part of the entropy that is due to
slow degrees of freedom; the contribution $S_{\text{v}}$ from the fast degrees
is usually called the vibrational entropy, and is taken to be the same for GL
and SCL as a standard and reliable approximation. Thus,
\[
S^{\text{(c)}}\equiv S-S_{\text{v}}.
\]
We now find from (\ref{Relax_S_Part}) that%
\begin{equation}
S_{\text{SCL}}^{\text{(c)}}=S_{\text{GL}}^{\text{(c)}}+S_{\text{relax}%
},\label{Conf_Entropy_Relation}%
\end{equation}
which shows that even $S_{\text{GL}}^{\text{(c)}}$ decreases during relaxation
as a consequence of the second law. This general result
(\ref{Conf_Entropy_Relation})\ is contradicted by the computation result of
Mauro et al \cite{Mauro}, according to which their so-called configurational
entropy reaches $S_{\text{SCL}}^{\text{(c)}}$ from below. It is clear that
their formulation of the configurational entropy, and therefore, the entropy
obtained by adding $S_{\text{v}}$ to it violates the second law. Hence, their
entropy \emph{cannot} represent the entropy of a glass.


\begin{thebibliography}{9}                                                                                                %


\bibitem {Mauro}J.C. Mauro, P.K. Gupta, R.J. Loucks, and A.K. Varshneya, J.
Non-Cryst. Solids, \textbf{355}, (2009), 600.

\bibitem {Note0}The temperature here means the temperature $T_{0}$ of the
medium, the heat bath, and not the temperature $T(t)$ of the glass which, as
we will establish here, changes with time until it becomes equal to $T_{0}$
once equilibrium has been established.

\bibitem {Landau}L.D.\ Landau, E.M. Lifshitz, \textit{Statistical Physics},
Vol. 1, Third Edition, Pergamon Press, Oxford (1986).
\end{thebibliography}
\end{document}